\documentclass[prd,preprint,nofootinbib]{revtex4}
\usepackage{amsmath}
\usepackage{graphicx}
\usepackage{slashed}
\usepackage{url}

\renewcommand\today{\number\year-\ifnum\month<10 0\fi\number\month-\ifnum\day<10 0\fi\number\day}

\newcommand\MSbar{\ensuremath{{\overline{\text{MS}}}}}
\newcommand\e e
\DeclareMathOperator\Real{Re}
\renewcommand\Re\Real

\newcommand\fold\otimes

\begin{document}
\title{Choosing the Factorization Scale in Perturbative QCD}
\author{\textbf{Fabio Maltoni}}
\affiliation{Center for Particle Physics and Phenomenology (CP3), Universit\'e catholique de Louvain, B-1348 Louvain-la-Neuve, Belgium}
\author{\textbf{Thomas McElmurry}}
\affiliation{Department of Physics, University of Wisconsin, 1150 University Avenue, Madison, Wisconsin 53706}
\author{\textbf{Robert Putman}}
\author{\textbf{Scott Willenbrock}}
\affiliation{Department of Physics, University of Illinois at Urbana-Champaign, 1110 West Green Street, Urbana, Illinois 61801}
\begin{abstract}
We define the collinear factorization scheme, which absorbs only the collinear physics into the parton distribution functions.
In order to isolate the collinear physics, we introduce a procedure to combine real and virtual corrections, canceling infrared singularities prior to integration.
In the collinear scheme, the factorization scale $\mu$ has a simple physical interpretation as a collinear cutoff.
We present a method for choosing the factorization scale and apply it to the Drell-Yan process; we find $\mu \approx Q/2$, where $Q$ is the vector-boson invariant mass.
We show that, for a wide variety of collision energies and $Q$, the radiative corrections are small in the collinear scheme for this choice of factorization scale.
\end{abstract}
\begin{flushright}
\url{hep-ph/0703156}\\
CP3-07-10
\end{flushright}
\maketitle

\section{Introduction} \label{s:intro}
In the parton model, a hadron is regarded as a collection of quarks, antiquarks, and gluons, each of which carries some fraction $x$ of the hadron's momentum, with a number density $f(x,\mu)$, where $\mu$ is the factorization scale.
Qualitatively, the factorization scale corresponds to the resolution with which the hadron is being probed.
To calculate the cross section for processes in hadron-hadron or lepton-hadron collisions, the partonic cross section is convolved with the corresponding parton distribution functions $f(x,\mu)$.
The partonic (hard-scattering) cross section is independent of the factorization scale $\mu$ at leading order in perturbative QCD, but depends logarithmically on $\mu$ at next-to-leading order and higher.
When calculated to all orders in perturbative QCD, the hadronic cross section is independent of $\mu$.
However, at any finite order in perturbation theory, the calculated hadronic cross section depends on $\mu$.
This dependence is usually significant at low orders in perturbation theory.
One could conclude that the only way to obtain a reliable prediction is to calculate higher-order corrections until the factorization-scale dependence is reduced.
However, there are many processes for which higher-order corrections are not available.
It is in these cases that the question of factorization-scale choice is most important.

If the hard-scattering cross section is characterized by a single scale $Q$ (such as the invariant mass of the lepton pair in Drell-Yan production), then the factorization scale is usually chosen to be of order $Q$, simply because there is no other scale in the problem.
However, this reveals only the order of magnitude of $\mu$.
It is common to vary the factorization scale over some interval, perhaps $Q/2\leq\mu\leq2Q$, but there is no objective argument for either the central value or the range of variation.
If the hard-scattering cross section depends on more than one scale, then the choice of factorization scale becomes an even murkier issue.

One argument against trying to do better than simply choosing $\mu\sim Q$ (we use $\sim$ to denote order-of-magnitude equality) goes as follows.
Since the hadronic cross section is a physical quantity that does not depend on any factorization scale, the factorization scale is unphysical, and therefore one cannot make a physical argument for its choice.
It is in part due to arguments such as this that there has been little effort to try to do better than choosing $\mu\sim Q$.%
\footnote{One attempt is based on the principle of minimal sensitivity \cite{Stevenson:1986cu}, and another on complete renormalization group improvement \cite{Maxwell:2000mm}.}

On the other hand, the statement is often made that the factorization scale separates the short-distance physics of the hard-scattering cross section from the long-distance hadronic physics \cite{Brock:1993sz}.
The qualitative statement made earlier, that $\mu$ corresponds to the resolution with which the hadron is probed, falls into this class.
This is contrary to the attitude that the factorization scale is unphysical.

In this paper we put forward a physical argument for the choice of factorization scale.
We restrict our attention to Drell-Yan production, but the argument can and hopefully will be extended to other processes in the future.

The physical argument we advance is that the factorization scale should be chosen such that collinear (long-distance) physics is included in the parton distribution functions, and noncollinear (short-distance) physics in the hard-scattering cross section.
This is not a new idea, and goes back to the origins of the parton model.%
\footnote{See, for example, Ref.~\cite{Collins:1990bu}.}
What is new is the implementation of this idea in practice.
We introduce a new factorization scheme, which we dub the collinear scheme, in which the factorization scale corresponds to a collinear cutoff.
We then argue for a method to choose the factorization scale in that scheme.

The method we pursue here was first proposed in the context of Higgs-boson production in association with bottom quarks \cite{Plehn:2002vy, Boos:2003yi}, and was refined and elaborated upon in Ref.~\cite{Maltoni:2003pn}.
An alternative approach in the same spirit has been developed in Ref.~\cite{Alwall:2004xw}.

All of the above studies are in the context of Higgs-boson (both charged and neutral) production via a bottom-quark distribution function, which is a rather exotic process, both in the initial and the final state.
However, the ideas developed there are of general validity, and should be applicable to all parton-model calculations.
In this paper we further develop the method of Ref.~\cite{Maltoni:2003pn} and apply it to one of the most basic processes of perturbative QCD, namely Drell-Yan production \cite{Kubar-Andre:1978uy,Altarelli:1979ub}.
Our hope is that this method can be generalized to all parton-model calculations, finally satisfying the desire for a systematic method of factorization-scale choice.

\section{The collinear factorization scheme} \label{s:coll}
The parton distribution functions are evolved from one factorization scale to another via the Dokshitzer-Gribov-Lipatov-Altarelli-Parisi (DGLAP) equations.
These equations sum collinear logarithms into the parton distribution functions.
In this section we define a factorization scheme in which the factorization scale $\mu$ has the interpretation of a cutoff in the integration over the virtuality of a propagator associated with collinear radiation.

\begin{figure}[htbp]
\begin{center}\includegraphics{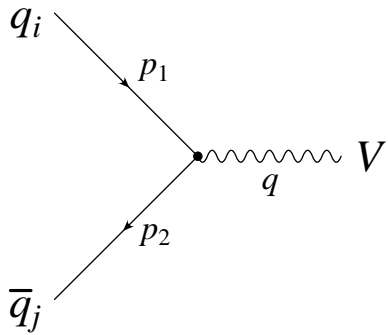}\end{center}
\caption{Drell-Yan production of an electroweak gauge boson $V$.}\label{f:qqx_V}
\end{figure}

Consider the Drell-Yan process, shown in Fig.~\ref{f:qqx_V}.
The colliding quark $q_i$ and antiquark $\bar q_j$ annihilate into an electroweak gauge boson $V$ of invariant mass $Q$; we do not consider the subsequent decay of the boson into lepton pairs, which is irrelevant to our discussion.
The leading-order (LO) cross section is
\begin{equation}
\sigma_{q\bar q}^{(0)}=\frac{4\pi^2\alpha}{3S}\sum_{i,j}C_{ij}(q_{0i}\fold\bar q_{0j}+\bar q_{0j}\fold q_{0i})(z_0),
\end{equation}
where $\sqrt S$ is the hadronic center-of-mass energy, $z_0\equiv Q^2/S$, $C_{ij}$ (shown in Table~\ref{t:couplings}) specifies the coupling of the quarks to the boson $V$, and the convolution $f_1\fold f_2$ of two functions $f_1$ and $f_2$ is defined by
\begin{equation}
(f_1\fold f_2)(x)\equiv\int_0^1dx_1\int_0^1dx_2\,\delta(x_1x_2-x)f_1(x_1)f_2(x_2).
\end{equation}

We will also need this cross section calculated in dimensional regularization with $4-2\epsilon$ spacetime dimensions ($\mu_D$ is the 't Hooft mass):
\begin{equation}\label{eq:sigma_LO_DR}
\sigma_{q\bar q}^{(0)}=\frac{4\pi^2\alpha}{3S}(1-\epsilon)\mu_D^{2\epsilon}\sum_{i,j}C_{ij}(q_{0i}\fold\bar q_{0j}+\bar q_{0j}\fold q_{0i})(z_0).
\end{equation}

\begin{table}[htbp]
\caption{Couplings of the quarks to electroweak gauge bosons.
Here $s_W=\sin\theta_W$, $c_W=\cos\theta_W$, and $V_{ij}$ denotes an element of the Cabibbo-Kobayashi-Maskawa matrix.}\label{t:couplings}
\begin{center}\begin{tabular}{cc}
\hline\hline
$q_i\bar q_j\to V$&$C_{ij}$\\
\hline
$u_i\bar u_j\to\gamma^*$&$\frac49\delta_{ij}$\\
$d_i\bar d_j\to\gamma^*$&$\frac19\delta_{ij}$\\
$u_i\bar d_j\to W^+$&$\frac1{4s_W^2}|V_{ij}|^2$\\
$d_i\bar u_j\to W^-$&$\frac1{4s_W^2}|V_{ji}|^2$\\
$u_i\bar u_j\to Z$&$\frac1{8s_W^2c_W^2}(1-\frac83s_W^2+\frac{32}9s_W^4)\delta_{ij}$\\
$d_i\bar d_j\to Z$&$\frac1{8s_W^2c_W^2}(1-\frac43s_W^2+\frac89s_W^4)\delta_{ij}$\\
\hline
\end{tabular}\end{center}
\end{table}

\subsection{Initial gluons}\label{ss:init}
\begin{figure}[htbp]
\begin{center}\includegraphics{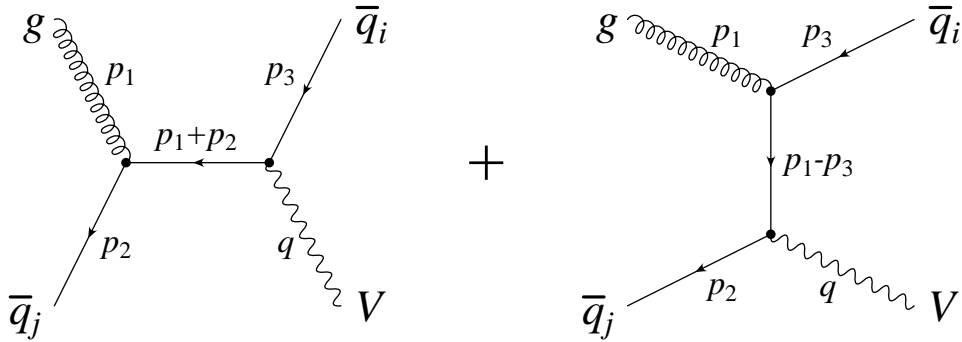}\end{center}
\caption{Correction to the production of an electroweak gauge boson $V$ due to initial gluons.}\label{f:gqx_qxV}
\end{figure}

The quark distribution function $q_i(x)$ receives corrections due to the splitting of a parton into a pair of collinear partons, one of which is a quark.
In order to find the correction to $q_i(x)$ arising from the splitting $g\to\bar q_iq_i$, we consider the next-to-leading-order (NLO) correction to the Drell-Yan cross section due to the process $g\bar q_j\to\bar q_iV$, shown in Fig.~\ref{f:gqx_qxV}.

Let $s\equiv(p_1+p_2)^2$, $t\equiv(p_1-p_3)^2$, and $u\equiv(p_2-p_3)^2$ be the usual Mandelstam variables, with $s+t+u=Q^2$.
The cross section for this process is
\begin{multline}
\sigma_{g\bar q}^{(1)}=\frac{\pi\alpha\alpha_s}{3S}\sum_{i,j}C_{ij}\int_{Q^2}^S\frac{ds}{s^2}\int_{-\infty}^0dt\int_{-\infty}^0du\,\delta(s+t+u-Q^2)\\
\times(g\fold\bar q_j+\bar q_j\fold g)\left(\frac sS\right)\left[\frac s{-t}+\frac{-t}s-\frac{2uQ^2}{st}\right].
\end{multline}
This expression manifestly shows the singular behavior near $t=0$, which corresponds to the splitting of the initial gluon into a collinear quark-antiquark pair.
The integrand contains a simple pole at $t=0$; thus this expression for the cross section is divergent.

\begin{figure}[htbp]
\begin{center}\includegraphics[scale=1.7]{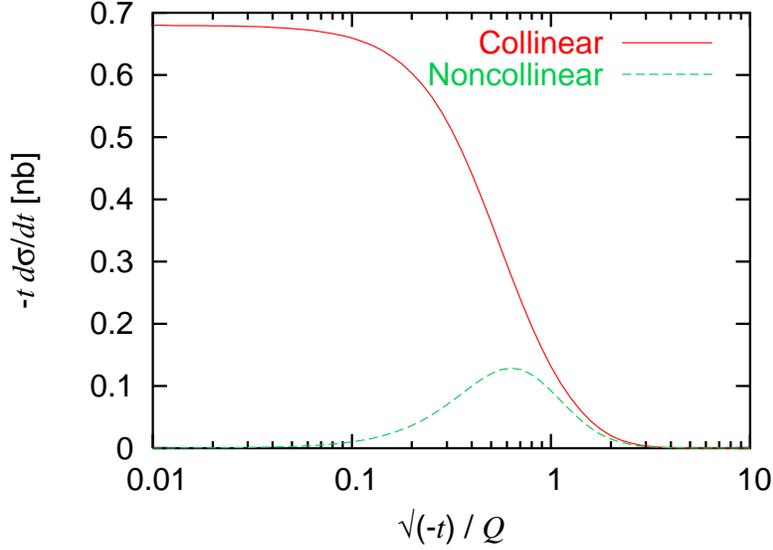}\end{center}
\caption{ The quantities $-t\,d\sigma_{g\bar q+qg}^{(1,\text{col})}/dt$ and $-t\,d\sigma_{g\bar q+qg}^{(1,\text{non})}/dt$, defined via Eqs.~(\ref{eq:sigma_init_c}) and (\ref{eq:sigma_init_nc}), for the case of real $Z$-boson production at the Tevatron.
The ``collinear'' curve, $-t\,d\sigma_{g\bar q+qg}^{(1,\text{col})}/dt$, passes through 50\% of its limiting value when $\sqrt{-t}=0.53Q$.}\label{f:ncdZ_0091TeV_init}
\end{figure}

If we use the delta function to perform the integration over $u$, we can break the cross section into two pieces: a ``collinear'' piece, comprising the terms proportional to $1/t$, and a ``noncollinear'' piece, comprising all other terms.
Thus $\sigma_{g\bar q}^{(1)}=\sigma_{g\bar q}^{(1,\text{col})}+\sigma_{g\bar q}^{(1,\text{non})}$, where
\begin{equation}\label{eq:sigma_init_c}
\sigma_{g\bar q}^{(1,\text{col})}=\frac{2\pi\alpha\alpha_s}{3S}\sum_{i,j}C_{ij}\int_{Q^2}^S\frac{ds}s\int_{-s+Q^2}^0\frac{dt}{-t}\,(g\fold\bar q_j+\bar q_j\fold g)\left(\frac sS\right)P_{qg}\left(\frac{Q^2}s\right),
\end{equation}
\begin{equation}\label{eq:sigma_init_nc}
\sigma_{g\bar q}^{(1,\text{non})}=\frac{\pi\alpha\alpha_s}{3S}\sum_{i,j}C_{ij}\int_{Q^2}^S\frac{ds}{s^2}\int_{-s+Q^2}^0dt\,(g\fold\bar q_j+\bar q_j\fold g)\left(\frac sS\right)\frac{2Q^2-t}s,
\end{equation}
and $P_{qg}(z)=\frac12\big[z^2+(1-z)^2\big]$ is the DGLAP splitting function.
The differential cross sections $-t\,d\sigma_{g\bar q}^{(1,\text{col})}/dt$ and $-t\,d\sigma_{g\bar q}^{(1,\text{non})}/dt$ corresponding to these two pieces (combined with their counterparts corresponding to the analogous process $q_ig\to Vq_i$) are shown, for the case of real $Z$-boson production at the Tevatron, in Fig.~\ref{f:ncdZ_0091TeV_init}.\footnote{%
We take the renormalization scale for $\alpha_s$ to be $Q$, here and throughout this paper.
This is a separate issue from the choice of the factorization scale.}
The two curves resemble a ``plateau'' and a ``hump,'' respectively, and their contributions to the cross section are proportional to the areas under the curves.
The noncollinear contribution is small, while the collinear divergence is evident in the plateau, which extends infinitely far to the left.
We must absorb this divergence into the parton distribution functions if we are to obtain a finite prediction for the cross section.

In order to do this, we define the collinear factorization scheme as follows.
We define a counterterm $\bar\sigma_{g\bar q}^{(1)}$ by imposing a cutoff $-\mu^2$ on the integration over $t$ and replacing the integrand with $1/t$ times the residue of the collinear pole:
\begin{align}
\bar\sigma_{g\bar q}^{(1)}&\equiv\left[\lim_{t\to0}\left(-t\frac{d\sigma_{g\bar q}^{(1)}}{dt}\right)\right]\int_{-\mu^2}^0\frac{dt}{-t}\\
&=\frac{2\pi\alpha\alpha_s}{3S}\sum_{i,j}C_{ij}\int_{Q^2}^S\frac{ds}s\,(g\fold\bar q_j+\bar q_j\fold g)\left(\frac sS\right)P_{qg}\left(\frac{Q^2}s\right)\int_{-\mu^2}^0\frac{dt}{-t}.\label{eq:sigma_init_ct}
\end{align}
The counterterm is determined solely by the collinear part of the cross section; Eq.~(\ref{eq:sigma_init_ct}) is identical to Eq.~(\ref{eq:sigma_init_c}) except for the replacement $s-Q^2\to\mu^2$ in the lower limit of the $t$-integral.
We regard $\mu$ as the factorization scale in this scheme.
By subtracting the counterterm from the bare cross section, we obtain the explicit correction in the collinear scheme:
\begin{multline}\label{eq:sigma_init_exp_raw}
\sigma_{g\bar q}^{(1)}-\bar\sigma_{g\bar q}^{(1)}=\frac{2\pi\alpha\alpha_s}{3S}\sum_{i,j}C_{ij}\int_{Q^2}^S\frac{ds}s\,(g\fold\bar q_j+\bar q_j\fold g)\left(\frac sS\right)\\
\times\left[\int_{-s+Q^2}^0dt\,\left(\frac1{-2t}+\frac{-t}{2s^2}-\frac{Q^2(s-Q^2+t)}{s^2(-t)}\right)-P_{qg}\left(\frac{Q^2}s\right)\int_{-\mu^2}^0\frac{dt}{-t}\right].
\end{multline}
By a rearrangement of terms,
\begin{multline}\label{eq:sigma_init_exp_regroup}
\sigma_{g\bar q}^{(1)}-\bar\sigma_{g\bar q}^{(1)}=\frac{2\pi\alpha\alpha_s}{3S}\sum_{i,j}C_{ij}\int_{Q^2}^S\frac{ds}s\,(g\fold\bar q_j+\bar q_j\fold g)\left(\frac sS\right)\\
\times\left[\int_{-s+Q^2}^{-\mu^2}\frac{dt}{-t}\,P_{qg}\left(\frac{Q^2}s\right)+\int_{-s+Q^2}^0dt\,\frac{2Q^2-t}{2s^2}\right],
\end{multline}
we obtain an expression in which the integrals are finite.
The result is
\begin{multline}\label{eq:sigma_init_exp}
\sigma_{g\bar q}^{(1)}-\bar\sigma_{g\bar q}^{(1)}=\frac{2\pi\alpha\alpha_s}{3S}\sum_{i,j}C_{ij}\int_{z_0}^1\frac{dz}z\,(g\fold\bar q_j+\bar q_j\fold g)\left(\frac{z_0}z\right)\\
\times\left[P_{qg}(z)\ln\frac{Q^2(1-z)}{\mu^2z}+\frac14(1+2z-3z^2)\right].
\end{multline}

If an explicit expression for the counterterm $\bar\sigma_{g\bar q}^{(1)}$ is desired, it is necessary to regulate the collinear divergence.
This can be done by working in dimensional regularization and suitably generalizing the definition of the counterterm:
\begin{equation}
\bar\sigma_{g\bar q}^{(1)}\equiv\left[\lim_{t\to0}\left((-t)^{1+\epsilon}\frac{d\sigma_{g\bar q}^{(1)}}{dt}\right)\right]\int_{-\mu^2}^0\frac{dt}{(-t)^{1+\epsilon}}.
\end{equation}
We perform the integration over $t$ and expand in powers of $\epsilon$ to obtain
\begin{multline}
\bar\sigma_{g\bar q}^{(1)}=-\frac{2\pi\alpha\alpha_s}{3S}(1-\epsilon)\mu_D^{2\epsilon}\sum_{i,j}C_{ij}\int_{z_0}^1\frac{dz}z\,(g\fold\bar q_j+\bar q_j\fold g)\left(\frac{z_0}z\right)\\
\times\left[\left(\frac1\epsilon-\gamma+\ln\frac{4\pi\mu_D^2}{\mu^2(1-z)}\right)P_{qg}(z)-z(1-z)\right].
\end{multline}
This result can be expressed concisely as
\begin{equation}
\bar\sigma_{g\bar q}^{(1)}=\frac{4\pi^2\alpha}{3S}(1-\epsilon)\mu_D^{2\epsilon}\sum_{i,j}C_{ij}(\delta q_i\fold\bar q_j+\bar q_j\fold\delta q_i)(z_0),
\end{equation}
where
\begin{equation}
\delta q_i(x)=-\frac{\alpha_s}{2\pi}\int_x^1\frac{dz}z\,g\left(\frac xz\right)\left[\left(\frac1\epsilon-\gamma+\ln\frac{4\pi\mu_D^2}{\mu^2(1-z)}\right)P_{qg}(z)-z(1-z)\right].
\end{equation}
The correction $\delta\bar q_j(x)$ to the antiquark distribution function, due to the process $q_ig\to Vq_j$, is identical.
These corrections are absorbed into the parton distribution functions if we define $q_i(x)=q_{0i}(x)+\delta q_i(x)$ and $\bar q_j(x)=\bar q_{0j}(x)+\delta\bar q_j(x)$, and calculate the LO cross section, Eq.~(\ref{eq:sigma_LO_DR}), with $q_i(x)$ and $\bar q_j(x)$ in place of $q_{0i}(x)$ and $\bar q_{0j}(x)$.
The explicit correction, $\sigma_{g\bar q}^{(1)}-\bar\sigma_{g\bar q}^{(1)}$, calculated in dimensional regularization, matches Eq.~(\ref{eq:sigma_init_exp}).

We come now to the question of the choice of $\mu$.
It is possible to choose $\mu$ such that $\bar\sigma_{g\bar q}^{(1)}=\sigma_{g\bar q}^{(1,\text{col})}$.
With this choice, the collinear physics is entirely absorbed into the parton distribution functions, and the explicit correction to the cross section is equal to $\sigma_{g\bar q}^{(1,\text{non})}$.
However, a precise determination of the value of $\mu$ that accomplishes this is unnecessary; in practice one can come close by choosing $\mu$ to equal the value of $\sqrt{-t}$ for which $-t\,d\sigma_{g\bar q}^{(1,\text{col})}/dt$ passes through 50\% of its limiting value.
This is the convention we will follow in this paper.
For the case of real $Z$-boson production at the Tevatron (Fig.~\ref{f:ncdZ_0091TeV_init}), this prescription indicates a scale $\mu=0.53Q$.

\begin{figure}[htbp]
\begin{center}\includegraphics[scale=1.7]{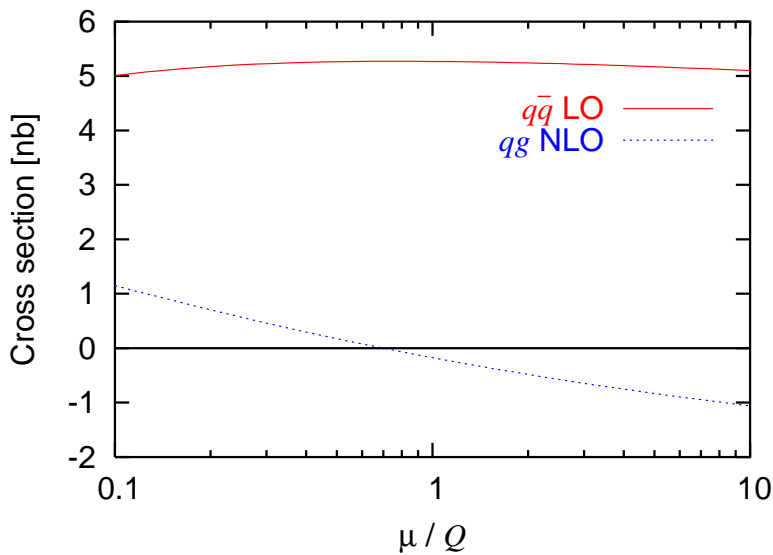}\end{center}
\caption{The factorization-scale dependence in the collinear scheme of the LO and initial-gluon NLO contributions to real $Z$-boson production at the Tevatron.
The factorization scale indicated by the plateau is $0.53Q$.}\label{f:tZ_0091TeV_init}
\end{figure}

Fig.~\ref{f:tZ_0091TeV_init} shows the factorization-scale dependence of the explicit correction to the cross section in the collinear scheme, given by Eq.~(\ref{eq:sigma_init_exp}) (together with the analogous contribution from $q_ig\to Vq_j$), for the case of real $Z$-boson production at the Tevatron.
This correction is small near the scale $\mu=0.53Q$, which supports our argument for this scale.
However, the correction shown in Fig.~\ref{f:tZ_0091TeV_init} is also small for $\mu=Q$, so we have not yet demonstrated the superiority of using a scale other than $Q$.
This will become evident when we consider higher values of $Q$ in Section~\ref{s:results}.

\subsection{Real and virtual gluons}
\begin{figure}[htbp]
\begin{center}\includegraphics{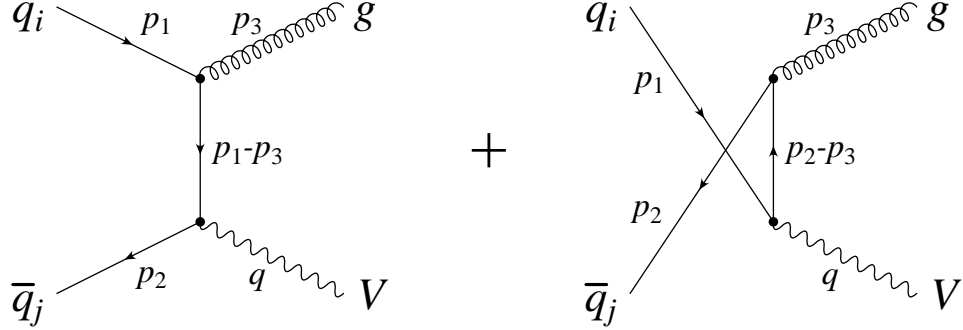}\end{center}
\caption{Correction to the production of an electroweak gauge boson $V$ due to real gluon emission.}\label{f:qqx_gV}
\end{figure}

\begin{figure}[htbp]
\begin{center}\includegraphics{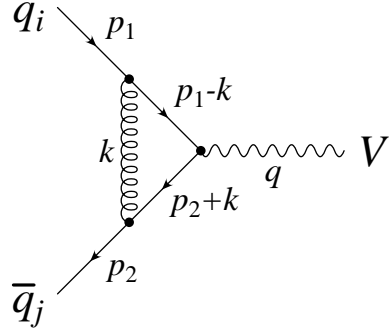}\end{center}
\caption{Vertex correction to the production of an electroweak gauge boson $V$.}\label{f:qqx_V_vtx}
\end{figure}

Now consider the NLO correction due to gluon radiation, shown in Fig.~\ref{f:qqx_gV}.
The cross section is
\begin{multline}\label{eq:sigma_real}
\sigma_{q\bar q}^{(1,\text{real})}=\frac{8\pi\alpha\alpha_S}{9S}\sum_{i,j}C_{ij}\int_{Q^2}^\infty\frac{ds}{s^2}\int_{-\infty}^0dt\int_{-\infty}^0du\,\delta(s+t+u-Q^2)\\
\times(q_i\fold\bar q_j+\bar q_j\fold q_i)\left(\frac sS\right)\left[\frac tu+\frac ut+\frac{2sQ^2}{tu}\right].
\end{multline}
Using the identity
\begin{equation}
\frac1{tu}=\frac1{s-Q^2}\left(\frac1{-t}+\frac1{-u}\right),
\end{equation}
we rewrite this as
\begin{multline}
\sigma_{q\bar q}^{(1,\text{real})}=\frac{8\pi\alpha\alpha_s}{9S}\sum_{i,j}C_{ij}\int_{Q^2}^\infty\frac{ds}{s^2}\int_{-\infty}^0dt\int_{-\infty}^0du\,\delta(s+t+u-Q^2)\\
\times(q_i\fold\bar q_j+\bar q_j\fold q_i)\left(\frac sS\right)\left[\frac{s^2+Q^4}{s-Q^2}\left(\frac1{-t}+\frac1{-u}\right)-2\right].
\end{multline}
There are two collinear singularities, at $t=0$ and at $u=0$, corresponding to the emission of a collinear gluon by the initial quark and antiquark, respectively.
Each is accompanied by an infrared divergence $1/(s-Q^2)$.
This soft singularity resides at the intersection of the two collinear singularities ($t=u=0$) and threatens to obscure the collinear physics in which we are interested.
It is cancelled, however, by the infrared divergence from diagrams involving virtual gluons.
We therefore turn our attention to these, beginning with the vertex correction shown in Fig.~\ref{f:qqx_V_vtx}.

The inclusion of this diagram modifies the tree-level amplitude by the replacement $\gamma^\mu\to\gamma^\mu\big[1+\delta F_1(Q^2)\big]$, where
\begin{equation}
\gamma^\mu\delta F_1(Q^2)=-\frac{16}3\pi i\alpha_s\int\frac{d^4k}{(2\pi)^4}\,\left[\frac{\gamma_\nu(\slashed k+\slashed p_2)\gamma^\mu(\slashed k-\slashed p_1)\gamma^\nu}{k^2(k-p_1)^2(k+p_2)^2}-\frac{\gamma_\nu(\slashed k+\slashed p_2)\gamma^\mu(\slashed k-\slashed p_1)\gamma^\nu}{(k^2-\Lambda^2)(k-p_1)^2(k+p_2)^2}\right].
\end{equation}
In the second term we have introduced the Pauli-Villars regulator $\Lambda\gg Q$.
The loop integration is simplified by introducing Feynman parameters $x,y,z$ and shifting the loop momentum to $\ell\equiv k-xp_1+yp_2$.
The result, up to terms that vanish in the $\Lambda\to\infty$ limit, is \cite{Peskin:1995ev}
\begin{align}
\delta F_1(Q^2)&=\begin{aligned}[t]
&-\frac{32}3\pi i\alpha_s\int_0^1dx\int_0^1dy\int_0^1dz\,\delta(x+y+z-1)\\
&\times\int\frac{d^4\ell}{(2\pi)^4}\,\left[\frac{\ell^2-2(z+xy)Q^2}{(\ell^2+xyQ^2)^3}-\frac{\ell^2}{(\ell^2-z\Lambda^2)^3}\right]
\end{aligned}\\
&=\begin{aligned}[t]
&-\frac{2\alpha_s}{3\pi}\int_0^1dx\int_0^1dy\int_0^1dz\,\delta(x+y+z-1)\\
&\times\left[\frac z{1-z}\left(\frac1x+\frac1y\right)+1-\ln\left(-\frac{z\Lambda^2}{xyQ^2}\right)\right].
\end{aligned}
\end{align}
The integrand contains terms proportional to $1/x$ and $1/y$, which correspond to collinear singularities at $x=0$ and at $y=0$.
Each is accompanied by a factor $1/(1-z)$; this is the infrared singularity which cancels against that of the real-gluon contribution.
The term involving $\ln\Lambda$ is an ultraviolet divergence.

\begin{figure}[htbp]
\begin{center}\includegraphics{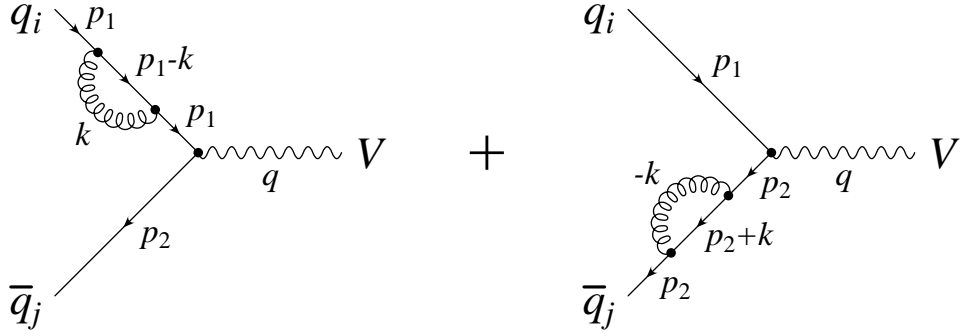}\end{center}
\caption{Wavefunction renormalization correction to the production of an electroweak gauge boson $V$.}\label{f:qqx_V_wfr}
\end{figure}

We must also include the correction due to wavefunction renormalization of the external quark lines, shown in Fig.~\ref{f:qqx_V_wfr}.
This involves the quark self-energy $\Sigma(p)$, which for massless quarks takes the form $\Sigma(p)=\Sigma'(p^2)\slashed p$.
The tree-level amplitude is  modified by the multiplicative factor $1/\sqrt{1+\Sigma'(0)}$ for each external leg.

Let us first consider the correction to the incoming quark.
The self-energy is
\begin{equation}
\Sigma(p_1)=\frac{32}3\pi i\alpha_s\int\frac{d^4k}{(2\pi)^4}\left[\frac{\slashed k-\slashed p_1}{k^2(k-p_1)^2}-\frac{\slashed k-\slashed p_1}{(k^2-\Lambda^2)(k-p_1)^2}\right].
\end{equation}
This integral contains an ultraviolet divergence (regulated by the inclusion of the second term), which cancels against that of the vertex correction.
It also contains a collinear divergence, which we want to express in a form that can be easily combined with those we have encountered in the vertex correction.
The natural way to proceed with the calculation of $\Sigma(p_1)$ is to introduce Feynman parameters $x$ and $y$, with $x+y=1$.
This will not suffice for our present purposes; we need a two-dimensional Feynman-parameter space in order to match the vertex correction.
We can achieve this by multiplying by unity:
\begin{equation}
\Sigma(p_1)=\frac{32}3\pi i\alpha_s\int\frac{d^4k}{(2\pi)^4}\left[\frac{(\slashed k-\slashed p_1)(k+p_2)^2}{k^2(k-p_1)^2(k+p_2)^2}-\frac{(\slashed k-\slashed p_1)(k+p_2)^2}{(k^2-\Lambda^2)(k-p_1)^2(k+p_2)^2}\right].
\end{equation}
The denominator now matches the denominator of the vertex correction, and we can therefore introduce matching Feynman parameters.

We can think of $p_2$, which has appeared out of left field, as the momentum of the incoming antiquark in $q_i\bar q_j\to V$.
We take the antiquark and the vector boson to be on shell, \textit{i.e.}~$p_2^2=0$ and $(p_1+p_2)^2=Q^2$, but $p_1$ must be kept off shell for the moment.

We introduce Feynman parameters $x,y,z$ and shift the loop momentum to $\ell\equiv k-xp_1+yp_2$:
\begin{multline}
\Sigma(p_1)=-\frac{32}3\pi i\alpha_s\int_0^1dx\int_0^1dy\int_0^1dz\,\delta(x+y+z-1)\\
\times\int\frac{d^4\ell}{(2\pi)^4}\left[\frac1{(\ell^2+xyQ^2+xzp_1^2)^3}-\frac1{(\ell^2-z\Lambda^2)^3}\right]\\
\times\bigg\{\big[(2-3x)\slashed p_1-(1-3y)\slashed p_2\big]\ell^2+2x\big[(1-x)\slashed p_1+y\slashed p_2\big]\big[(1-y)Q^2-zp_1^2\big]\bigg\}.
\end{multline}
The presence of terms proportional to $\slashed p_2$ may be surprising, but these terms integrate to zero and do not contain a collinear singularity.
Thus they need trouble us no further.
At this point we can read off the coefficient of $\slashed p_1$ and set $p_1^2=0$:
\begin{align}
\Sigma'(0)&=\begin{aligned}[t]
&-\frac{32}3\pi i\alpha_s\int_0^1dx\int_0^1dy\int_0^1dz\,\delta(x+y+z-1)\\
&\times\int\frac{d^4\ell}{(2\pi)^4}\,\left[\frac{(2-3x)\ell^2+2x(1-x)(1-y)Q^2}{(\ell^2+xyQ^2)^3}-\frac{(2-3x)\ell^2}{(\ell^2-z\Lambda^2)^3}\right]
\end{aligned}\\
&=\frac{2\alpha_s}{3\pi}\int_0^1dx\int_0^1dy\int_0^1dz\,\delta(x+y+z-1)\left[\frac zy+x+(2-3x)\ln\left(-\frac{z\Lambda^2}{xyQ^2}\right)\right].
\end{align}
The term involving $\ln\Lambda$ is the expected ultraviolet divergence.
There is also a single collinear singularity, at $y=0$.

The self-energy correction to the incoming antiquark line is completely analogous, but it is appropriate to interchange the parameters $x$ and $y$.
The order-$\alpha_s$  contribution to the $q_i\bar q_j\to V$ amplitude due to virtual gluons is therefore equal to the tree-level amplitude times the factor
\begin{align}
\Delta&\equiv\delta F_1(Q^2)-\frac12\Sigma'(0)-\frac12\Sigma'(0)\\
&=\begin{aligned}[t]
&-\frac{\alpha_s}{3\pi}\int_0^1dx\int_0^1dy\int_0^1dz\,\delta(x+y+z-1)\\
&\times\left[\frac{z(3-z)}{1-z}\left(\frac1x+\frac1y\right)+3-z-(1-3z)\ln\left(-\frac{z\Lambda^2}{xyQ^2}\right)\right].
\end{aligned}
\end{align}
This expression still contains the ultraviolet-divergent term $\ln(\Lambda^2/Q^2)$.
However, its coefficient integrates to zero, so that we can discard this term.
The NLO contribution to the cross section is then
\begin{align}
\sigma_{q\bar q}^{(1,\text{virt})}&=(2\Re\Delta)\sigma_{q\bar q}^{(0)}\\
&=\begin{aligned}[t]
&-\frac{8\pi\alpha\alpha_s}{9S}\sum_{i,j}C_{ij}(q_i\fold\bar q_j+\bar q_j\fold q_i)(z_0)\int_0^1dx\int_0^1dy\int_0^1dz\,\delta(x+y+z-1)\\
&\times\left[\frac{z(3-z)}{1-z}\left(\frac1x+\frac1y\right)+3-z-(1-3z)\ln\frac z{xy}\right].
\end{aligned}
\end{align}
This must be combined with the correction due to real-gluon emission to yield an infrared-finite result.

\begin{figure}[htbp]
\begin{center}\includegraphics{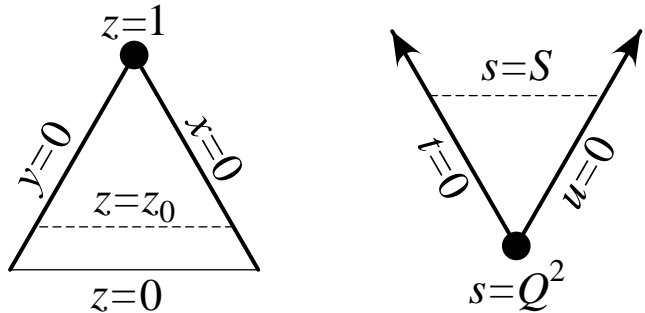}\end{center}
\caption{The region of Feynman-parameter space bounded by $x=0$, $y=0$, and $z=0$ is mapped by Eq.~(\ref{eq:Feynmandelstam}) onto the region of Mandelstam-variable space bounded by $s=\infty$, $t=0$, and $u=0$.
The real-gluon contribution is bounded by $s=S$ (or $z=z_0$), while the virtual-gluon contribution extends all the way to $z=0$ (or $s=\infty$).
The heavy lines represent the collinear singularities, and the black dots represent the infrared singularity.}\label{f:Feynmandelstam}
\end{figure}

The hadronic real-gluon-emission cross section involves an integration over the Mandelstam variables $s$, $t$, and $u$, subject to the constraint $s+t+u=Q^2$.
There are collinear singularities at $t=0$ and at $u=0$, and where the two meet there is an infrared singularity at $s=Q^2$.
Similarly, the virtual-gluon cross section involves an integration over the Feynman parameters $x$, $y$, and $z$, constrained by $x+y+z=1$.
There are collinear singularities at $x=0$ and at $y=0$, and where the two meet there is an infrared singularity at $z=1$.
The two spaces can be mapped onto one another in such a way that the singularity structures match exactly:
\begin{equation}\label{eq:Feynmandelstam}
x\to\frac{-u}s,\qquad y\to\frac{-t}s,\qquad z\to\frac{Q^2}s.
\end{equation}
This mapping is illustrated in Fig.~\ref{f:Feynmandelstam}.

We thus change variables from $x,y,z$ to $s,t,u$, using the relation
\begin{equation}
\int_0^1dx\int_0^1dy\int_0^1dz\,\delta(x+y+z-1)=Q^2\int_{Q^2}^\infty\frac{ds}{s^3}\int_{-\infty}^0dt\int_{-\infty}^0du\,\delta(s+t+u-Q^2).
\end{equation}
The virtual-gluon cross section becomes
\begin{multline}\label{eq:sigma_virt}
\sigma_{q\bar q}^{(1,\text{virt})}=-\frac{8\pi\alpha\alpha_s}{9S}\sum_{i,j}C_{ij}(q_i\fold\bar q_j+\bar q_j\fold q_i)(z_0)\int_{Q^2}^\infty\frac{ds}{s^2}\int_{-\infty}^0dt\int_{-\infty}^0du\,\delta(s+t+u-Q^2)\\
\times\frac{Q^2}s\left[\frac{Q^2(3s-Q^2)}{s-Q^2}\left(\frac1{-t}+\frac1{-u}\right)+\frac{3s-Q^2}s-\frac{s-3Q^2}s\ln\frac{sQ^2}{tu}\right].
\end{multline}
In this equation the symbols $s$, $t$, and $u$, while they have been designed to resemble Mandelstam variables, are not defined in terms of external momenta, but rather as particular combinations of Feynman parameters given by Eq.~(\ref{eq:Feynmandelstam}).
In particular $s\neq(p_1+p_2)^2=Q^2$.

We add Eqs.~(\ref{eq:sigma_real}) and (\ref{eq:sigma_virt}) to obtain the total cross section due to real and virtual gluons:
\begin{multline}
\sigma_{q\bar q}^{(1)}=\frac{8\pi\alpha\alpha_s}{9S}\sum_{i,j}C_{ij}\int_{Q^2}^\infty\frac{ds}{s^2}\int_{-\infty}^0dt\int_{-\infty}^0du\,\delta(s+t+u-Q^2)\\
\times\bigg\{\theta(S-s)(q_i\fold\bar q_j+\bar q_j\fold q_i)\left(\frac sS\right)\left[\frac{s^2+Q^4}{s-Q^2}\left(\frac1{-t}+\frac1{-u}\right)-2\right]\\
-(q_i\fold\bar q_j+\bar q_j\fold q_i)\left(\frac{Q^2}S\right)\frac{Q^2}s\left[\frac{Q^2(3s-Q^2)}{s-Q^2}\left(\frac1{-t}+\frac1{-u}\right)+\frac{3s-Q^2}s-\frac{s-3Q^2}s\ln\frac{sQ^2}{tu}\right]\bigg\}.
\end{multline}
The infrared cancellation is now manifest, as the residue of the $s=Q^2$ pole vanishes.
The cancellation of the infrared divergence prior to integration is reminiscent of the work of Refs.~\cite{Soper:1998ye,Soper:1999xk,Nagy:2003qn}.

\begin{figure}[htbp]
\begin{center}\includegraphics[scale=1.7]{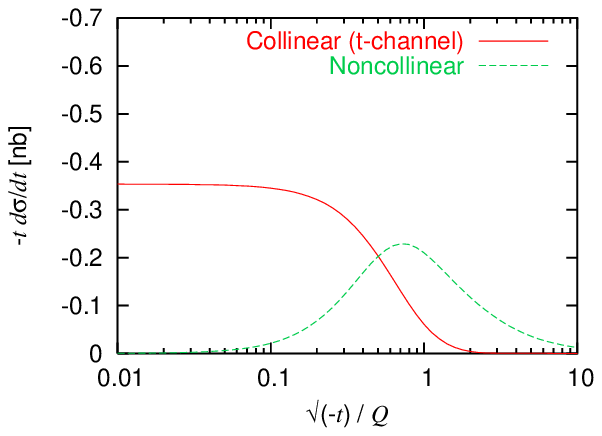}\end{center}
\caption{The quantities $-t\,d\sigma_{q\bar q}^{(1,\text{col},t)}/dt$ and $-t\,d\sigma_{q\bar q}^{(1,\text{non})}/dt$, defined via Eqs.~(\ref{eq:sigma_rv_c_t}) and (\ref{eq:sigma_rv_nc}), for the case of real $Z$-boson production at the Tevatron.
The ``collinear'' curve, $-t\,d\sigma_{q\bar q}^{(1,\text{col},t)}/dt$, passes through 50\% of its limiting value when $\sqrt{-t}=0.57Q$.}\label{f:ncdZ_0091TeV_rv}
\end{figure}

As in Eqs.~(\ref{eq:sigma_init_c}) and (\ref{eq:sigma_init_nc}), we decompose $\sigma_{q\bar q}^{(1)}$ into collinear and noncollinear pieces: $\sigma_{q\bar q}^{(1)}=\sigma_{q\bar q}^{(1,\text{col},t)}+\sigma_{q\bar q}^{(1,\text{col},u)}+\sigma_{q\bar q}^{(1,\text{non})}$, where
\begin{equation}\label{eq:sigma_rv_c_t}
\sigma_{q\bar q}^{(1,\text{col},t)}=\frac{8\pi\alpha\alpha_s}{9S}\sum_{i,j}C_{ij}\int_{Q^2}^\infty\frac{ds}s\int_{-s+Q^2}^0\frac{dt}{-t}\,\bigg[\begin{aligned}[t]
&\theta(S-s)(q_i\fold\bar q_j+\bar q_j\fold q_i)\left(\frac sS\right)\frac{s^2+Q^4}{s(s-Q^2)}\\
&-(q_i\fold\bar q_j+\bar q_j\fold q_i)\left(\frac{Q^2}S\right)\frac{Q^4(3s-Q^2)}{s^2(s-Q^2)}\bigg],
\end{aligned}
\end{equation}
\begin{multline}\label{eq:sigma_rv_nc}
\sigma_{q\bar q}^{(1,\text{non})}=\frac{8\pi\alpha\alpha_s}{9S}\sum_{i,j}C_{ij}\int_{Q^2}^\infty\frac{ds}{s^2}\int_{-\infty}^0dt\int_{-\infty}^0du\,\delta(s+t+u-Q^2)\\
\times\bigg\{\theta(S-s)(q_i\fold\bar q_j+\bar q_j\fold q_i)\left(\frac sS\right)[-2]\\
-(q_i\fold\bar q_j+\bar q_j\fold q_i)\left(\frac{Q^2}S\right)\frac{Q^2}s\left[\frac{3s-Q^2}s-\frac{s-3Q^2}s\ln\frac{sQ^2}{tu}\right]\bigg\},
\end{multline}
and $\sigma_{q\bar q}^{(1,\text{col},u)}$ is defined similarly to $\sigma_{q\bar q}^{(1,\text{col},t)}$.
The differential cross sections $-t\,d\sigma_{q\bar q}^{(1,\text{col},t)}/dt$ and $-t\,d\sigma_{q\bar q}^{(1,\text{non})}/dt$ are shown, for the case of real $Z$-boson production at the Tevatron, in Fig.~\ref{f:ncdZ_0091TeV_rv}.

The $t$-channel collinear-scheme counterterm takes the form
\begin{align}
\bar\sigma_{q\bar q}^{(1,t)}&=\frac{8\pi\alpha\alpha_s}{9S}\sum_{i,j}C_{ij}\int_{Q^2}^\infty\frac{ds}s\int_{-\mu^2}^0\frac{dt}{-t}\,\bigg[\begin{aligned}[t]
&\theta(S-s)(q_i\fold\bar q_j+\bar q_j\fold q_i)\left(\frac sS\right)\frac{s^2+Q^4}{s(s-Q^2)}\\
&-(q_i\fold\bar q_j+\bar q_j\fold q_i)\left(\frac{Q^2}S\right)\frac{Q^4(3s-Q^2)}{s^2(s-Q^2)}\bigg]
\end{aligned}\\
&=\frac{2\pi\alpha\alpha_s}{3S}\sum_{i,j}C_{ij}\int_{Q^2}^\infty\frac{ds}s\,(q_i\fold\bar q_j+\bar q_j\fold q_i)\left(\frac sS\right)P_{qq}\left(\frac{Q^2}s\right)\int_{-\mu^2}^0\frac{dt}{-t},\label{eq:sigma_rv_ct}
\end{align}
where $P_{qq}(z)=\frac43\big[(1+z^2)/(1-z)\big]_+$ is the DGLAP splitting function.
The $u$-channel counterterm $\bar\sigma_{q\bar q}^{(1,u)}$ is similar.
By subtracting these counterterms before integrating over $t$ and $u$, as in Eqs.~(\ref{eq:sigma_init_exp_raw}) and (\ref{eq:sigma_init_exp_regroup}), we arrive at the following result for the explicit correction in the collinear scheme:
\begin{multline}\label{eq:sigma_rv_exp}
\sigma_{q\bar q}^{(1)}-\bar\sigma_{q\bar q}^{(1,\text{col},t)}-\bar\sigma_{q\bar q}^{(1,\text{col},u)}=\frac{2\pi\alpha\alpha_s}{3S}\sum_{i,j}C_{ij}\int_{z_0}^1\frac{dz}z\,(q_i\fold\bar q_j+\bar q_j\fold q_i)\left(\frac{z_0}z\right)\\
\times\bigg[2P_{qq}(z)\ln\frac{Q^2}{\mu^2}+\frac83(1+z^2)\left(\frac{\ln(1-z)}{1-z}\right)_+-\left(\frac{8\pi^2}9+\frac43\right)\delta(1-z)\\
-\frac83\frac{1+z^2}{1-z}\ln z-\frac83(1-z)\bigg].
\end{multline}

\begin{figure}[htbp]
\begin{center}\includegraphics[scale=0.75]{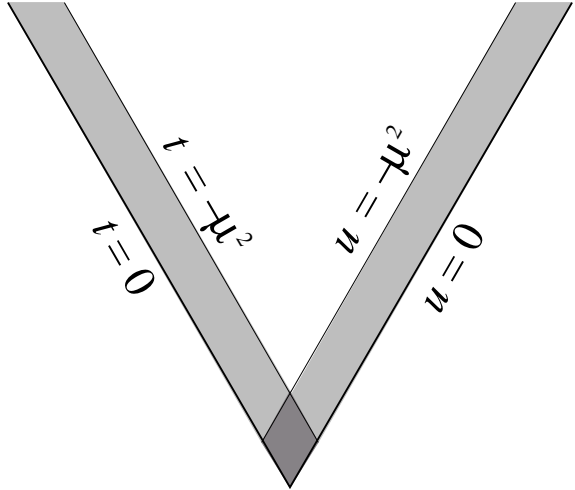}\end{center}
\caption{The counterterms corresponding to the $t$- and $u$-channel collinear singularities are formed by integrating over the shaded strips.
We must be careful to avoid double-counting in the region where the two strips overlap.}\label{f:ballpark}
\end{figure}

An explicit expression for the counterterms can be obtained by working in dimensional regularization:
\begin{multline}
\bar\sigma_{q\bar q}^{(1,t)}=\bar\sigma_{q\bar q}^{(1,u)}=-\frac{2\pi\alpha\alpha_s}{3S}(1-\epsilon)\mu_D^{2\epsilon}\sum_{i,j}C_{ij}\int_{z_0}^1\frac{dz}z\,(q_i\fold\bar q_j+\bar q_j\fold q_i)\left(\frac{z_0}z\right)\\
\times\bigg[\left(\frac1\epsilon-\gamma+\ln\frac{4\pi\mu_D^2}{\mu^2}\right)P_{qq}(z)-\frac43(1+z^2)\left(\frac{\ln(1-z)}{1-z}\right)_+\\
-\left(\frac{4\pi^2}3-\frac{14}3\right)\delta(1-z)-\frac43(1-z)\bigg].
\end{multline}
There is one added complication in dimensional regularization: it is necessary to correct for double counting in the region where both $t$ and $u$ are near zero (see Fig.~\ref{f:ballpark}).
Therefore, after subtracting the two counterterms, we must add the quantity
\begin{multline}
\bar{\bar\sigma}_{q\bar q}^{(1)}\equiv\left[\lim_{t,u\to0}\left((tu)^{1+\epsilon}\frac{d^2\sigma_{q\bar q}^{(1)}}{dt\,du}\right)\right]\int_{-\mu^2}^0dt\int_{-\mu^2}^0du\,\frac1{(tu)^{1+\epsilon}}\\
=\frac{2\pi\alpha\alpha_s}{3S}\sum_{i,j}C_{ij}\int_{z_0}^1\frac{dz}z\,(q_i\fold\bar q_j+\bar q_j\fold q_i)\left(\frac{z_0}z\right)\frac{8\pi^2}9\delta(1-z).
\end{multline}
By following this procedure, we recover the explicit correction given in Eq.~(\ref{eq:sigma_rv_exp}) \cite{McElmurry}.
The collinear contribution absorbed into the quark distribution function is
\begin{equation}
\bar\sigma_{q\bar q}^{(1,t)}-\frac12\bar{\bar\sigma}_{q\bar q}^{(1)}=\frac{4\pi^2\alpha}{3S}(1-\epsilon)\mu_D^{2\epsilon}\sum_{i,j}C_{ij}(\delta q_i\fold\bar q_j+\bar q_j\fold\delta q_i)\left(\frac{Q^2}S\right),
\end{equation}
where
\begin{multline}
\delta q_i(x)=-\frac{\alpha_s}{2\pi}\int_x^1\frac{dz}z\,q_i\left(\frac xz\right)\bigg[\left(\frac1\epsilon-\gamma+\ln\frac{4\pi\mu_D^2}{\mu^2}\right)P_{qq}(z)-\frac43(1+z^2)\left(\frac{\ln(1-z)}{1-z}\right)_+\\
-\left(\frac{8\pi^2}9-\frac{14}3\right)\delta(1-z)-\frac43(1-z)\bigg].
\end{multline}
The correction $\delta\bar q_j(x)$ to the antiquark distribution function is analogous.

\begin{figure}[htbp]
\begin{center}\includegraphics[scale=1.7]{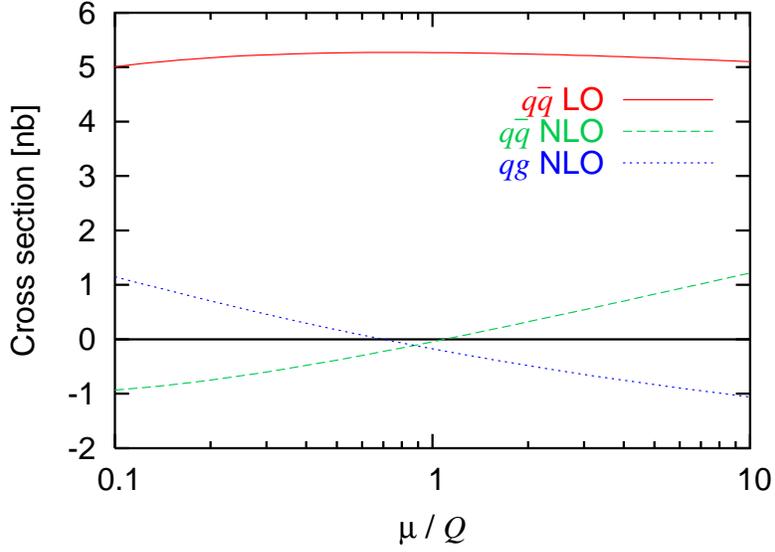}\end{center}
\caption{The factorization-scale dependence in the collinear scheme of the LO and NLO contributions to real $Z$-boson production at the Tevatron.
The factorization scales indicated by the plateaux are $0.53Q$ for initial gluons and $0.57Q$ for real and virtual gluons.}\label{f:tZ_0091TeV}
\end{figure}

The 50\% rule described in Section~\ref{ss:init}, applied to the plateau curve in Fig.~\ref{f:ncdZ_0091TeV_rv}, indicates a factorization scale $\mu=0.57Q$, about the same as that indicated by the initial-gluon plateau. 
Fig.~\ref{f:tZ_0091TeV} shows the factorization-scale dependence of both NLO corrections to the Drell-Yan cross section in the collinear scheme.
At $\mu\approx Q/2$, the correction due to real and virtual gluons is small, about $-7\%$ of the LO cross section.
This correction is even smaller at $\mu\approx Q$, essentially vanishing.
However, there is no reason to expect the correction to vanish at the appropriate factorization scale; it is enough that it is small, indicating a convergent perturbation series.

The collinear-scheme counterterms for the Drell-Yan process can be expressed simply in terms of convolutions of parton distribution functions with DGLAP splitting functions, as in Eqs.~(\ref{eq:sigma_init_ct}) and (\ref{eq:sigma_rv_ct}).
This suggests that the counterterms for other processes will have a similar form, and that NLO parton distribution functions defined in the collinear scheme will be universal.

\section{Results}\label{s:results}
In this section we show the factorization-scale dependence of the LO and NLO contributions to the Drell-Yan cross section, in the collinear scheme, for several values of $Q$ and $\sqrt S$.
In some cases the initial-gluon correction has been multiplied by a large factor in order to make it more visible.

\begin{figure}[htbp]
\begin{center}\includegraphics[scale=1.7]{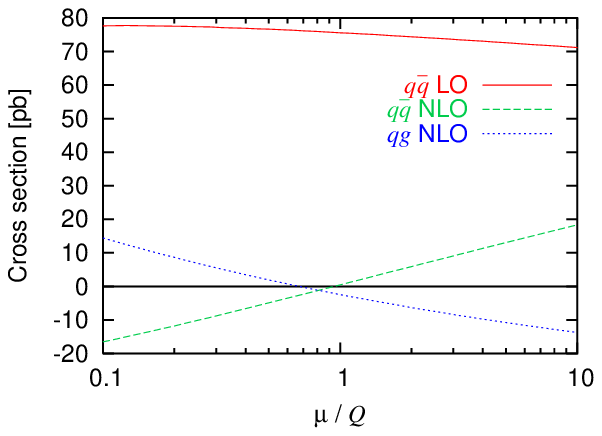}\end{center}
\caption{The factorization-scale dependence in the collinear scheme of the LO and NLO contributions to virtual $Z$-boson production at the LHC, with $Q=650~\text{GeV}$.
The factorization scales indicated by the plateaux are $0.52Q$ for initial gluons and $0.61Q$ for real and virtual gluons.}\label{f:tZ_0650LHC}
\end{figure}

Fig.~\ref{f:tZ_0650LHC} shows the case of virtual $Z$-boson production at the CERN Large Hadron Collider (LHC) ($\sqrt S=14~\text{TeV}$ $pp$), with $Q=650~\text{GeV}$.
This value of $Q$ has been chosen so that the ratio $Q/\sqrt S$ is the same as for the case of real $Z$-boson production at the Tevatron, presented in Fig~\ref{f:tZ_0091TeV}.
The factorization scales indicated by the plateaux are about the same as in that case, still in the vicinity of $\mu\approx Q/2$.
Again, both NLO corrections are small near this scale.

\begin{figure}[htbp]
\begin{center}\includegraphics[scale=1.7]{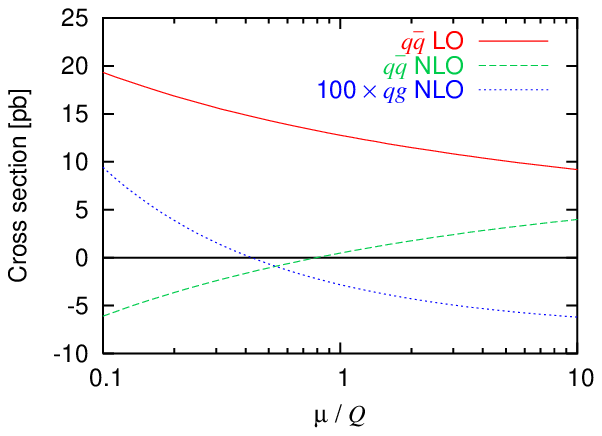}\end{center}
\caption{The factorization-scale dependence in the collinear scheme of the LO and NLO contributions to virtual $Z$-boson production at the Tevatron, with $Q=490\,\text{GeV}$.
The factorization scales indicated by the plateaux are $0.37Q$ for initial gluons and $0.63Q$ for real and virtual gluons.}\label{f:tZ_0490TeV}
\begin{center}\includegraphics[scale=1.7]{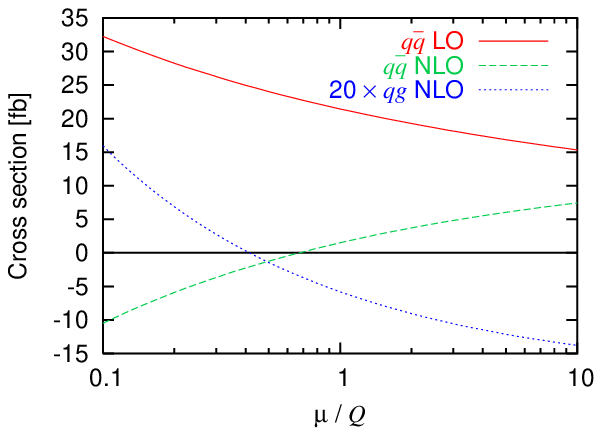}\end{center}
\caption{The factorization-scale dependence in the collinear scheme of the LO and NLO contributions to virtual $Z$-boson production at the LHC, with $Q=3.5\,\text{TeV}$.
The factorization scales indicated by the plateaux are $0.37Q$ for initial gluons and $0.58Q$ for real and virtual gluons.}\label{f:tZ_3500LHC}
\end{figure}

\begin{figure}[htbp]
\begin{center}\includegraphics[scale=1.7]{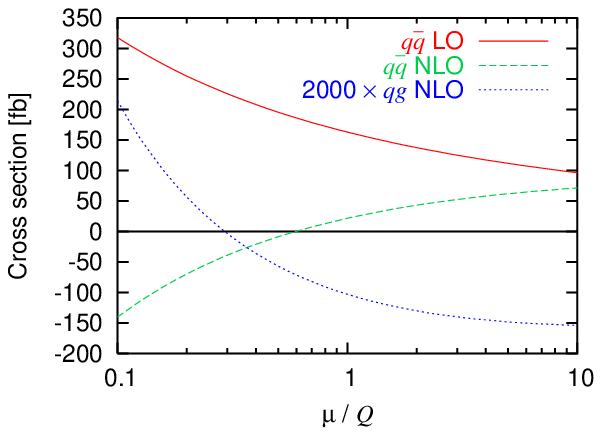}\end{center}
\caption{The factorization-scale dependence in the collinear scheme of the LO and NLO contributions to virtual $Z$-boson production at the Tevatron, with $Q=910\,\text{GeV}$.
The factorization scales indicated by the plateaux are $0.28Q$ for initial gluons and $0.53Q$ for real and virtual gluons.}\label{f:tZ_0910TeV}
\begin{center}\includegraphics[scale=1.7]{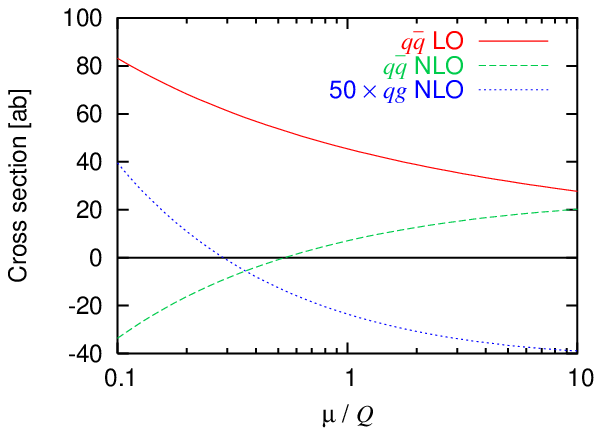}\end{center}
\caption{The factorization-scale dependence in the collinear scheme of the LO and NLO contributions to virtual $Z$-boson production at the LHC, with $Q=6.5\,\text{TeV}$.
The factorization scales indicated by the plateaux are $0.28Q$ for initial gluons and $0.49Q$ for real and virtual gluons.}\label{f:tZ_6500LHC}
\end{figure}

In Figs.~\ref{f:tZ_0490TeV}--\ref{f:tZ_6500LHC} we show the factorization-scale dependence of the virtual $Z$-boson production cross section for higher values of $Q$, at both the Tevatron and the LHC.
As $Q/\sqrt S$ increases, the factorization scales indicated by the plateaux decrease; this decrease is moderate for real and virtual gluons, and more pronounced for initial gluons.
As the scale indicated by the initial-gluon plateau decreases, we observe a corresponding decrease in the scale at which the initial-gluon correction is minimized, which supports our argument for this scale choice.
At large values of $Q/\sqrt S$, the scales indicated by the initial-gluon plateau and the real- and virtual-gluon plateau differ by nearly a factor of 2.
This is reflected in Figs.~\ref{f:tZ_0910TeV} and \ref{f:tZ_6500LHC}, where the initial-gluon correction is small for $\mu\approx Q/4$, while the real- and virtual-gluon correction is small for $\mu\approx Q/2$.
When choosing a single factorization scale, it is necessary to compromise between these two values.
Since the initial-gluon correction is dwarfed by the real- and virtual-gluon correction, for practical purposes one should choose the scale indicated by the real- and
virtual-gluon plateau, $\mu\approx Q/2$.

\begin{figure}[htbp]
\begin{center}\includegraphics[scale=1.6]{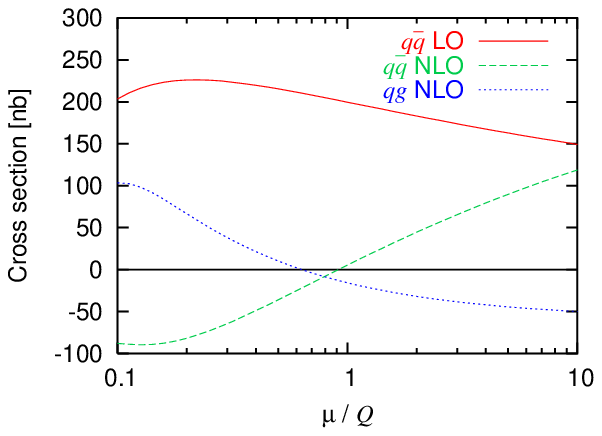}\end{center}
\caption{The factorization-scale dependence in the collinear scheme of the LO and NLO contributions to virtual photon production in $pp$ collisions at $\sqrt S=38.8\,\text{GeV}$, with $Q=5\,\text{GeV}$.
The factorization scales indicated by the plateaux are $0.48Q$ for initial gluons and $0.64Q$ for real and virtual gluons.}\label{f:tph0005usr}
\begin{center}\includegraphics[scale=1.6]{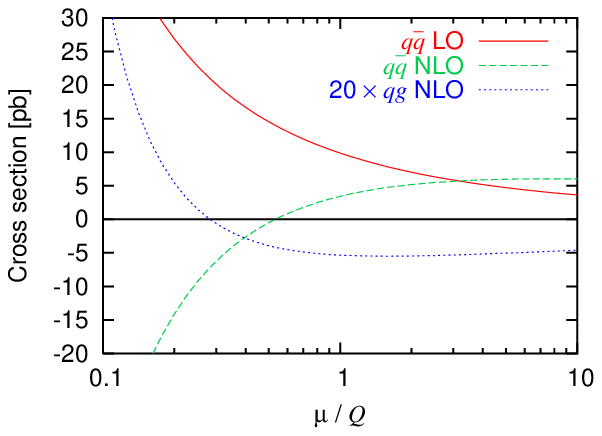}\end{center}
\caption{The factorization-scale dependence in the collinear scheme of the LO and NLO contributions to virtual photon production in $pp$ collisions at $\sqrt S=38.8\,\text{GeV}$, with $Q=20\,\text{GeV}$.
The factorization scales indicated by the plateaux are $0.27Q$ for initial gluons and $0.48Q$ for real and virtual gluons.}\label{f:tph0020usr}
\end{figure}

In Figs.~\ref{f:tph0005usr} and \ref{f:tph0020usr}, we examine Drell-Yan production of a virtual photon in a fixed-target experiment with a beam energy of $800~\text{GeV}$ ($\sqrt S=38.8~\text{GeV}$ $pp$).
Once again, the NLO corrections are small if the factorization scale is chosen according to the plateaux.
Fig.~\ref{f:tph0020usr} in particular illustrates the importance of this choice: if one instead chose $\mu=Q$, the NLO correction due to real and virtual gluons would equal 35\% of the LO cross section.

\section{Conclusion}
In this paper we have introduced the collinear factorization scheme, and argued for a method to choose the factorization scale in that scheme.
We applied this to the Drell-Yan process, and found that the explicit NLO corrections are small, less than 10\%, for a wide variety of machine energies and values of the vector-boson invariant mass $Q$.
The factorization scale indicated by our method is $\mu\approx Q/2$ in the collinear scheme.
This is not necessarily the correct scale in other factorization schemes, such as the popular \MSbar\ scheme.
We hope to adapt our method to choosing the factorization scale in the \MSbar\ scheme.

A complete calculation of a NLO cross section in the collinear scheme requires the LO partonic cross section to be convolved with NLO collinear-scheme parton distribution functions.%
\footnote{Evaluating the NLO corrections with NLO parton distribution functions is not necessary for formal consistency, as the difference is of order $\alpha_s^2$.}
Since these are not available at present, the numerical results presented in this paper have been computed using the LO CTEQ6L1 parton distribution functions \cite{Pumplin:2002vw}.
The total NLO correction then consists of the explicit radiative corrections plotted in Sections~\ref{s:coll} and \ref{s:results}, together with the implicit correction resulting from the replacement of LO parton distribution functions with NLO parton distribution functions in the evaluation of the LO cross section.
Without a NLO set of parton distribution functions in the collinear scheme, we cannot evaluate this implicit correction.
However, if the factorization scale is chosen in such a way that the explicit correction is small, any large correction must be contained in the implicit correction.
This suggests that a good approximation to the NLO cross section may be obtained by evaluating the LO cross section with NLO collinear-scheme parton distribution functions, with the factorization scale determined as above.
Such a set of NLO parton distribution functions could be generated once all the collinear-scheme counterterms are known.

\begin{acknowledgments}
We are grateful for conversations and correspondence with J.~Collins, D.~Soper, and G.~Sterman.
S.~W.~thanks the Aspen Center for Physics for hospitality.
This work was supported in part by the U.~S.~Department of Energy under contract No.~DE-FG02-91ER40677 and by the Belgian Federal Science Policy (IAP 6/11).
\end{acknowledgments}

\end{document}